\documentclass[a4paper,10pt]{article}
\usepackage{amsmath,amsthm,amssymb}
\usepackage{graphicx,subfigure}
\usepackage{url}

\usepackage{algorithm2e}
\usepackage{listings}
\usepackage{color}

\usepackage{multicol}
\def\twoplot[#1]#2#3#4#5{
\begin{figure}[h]
\begin{multicols}{2}
\begin{center}
    \includegraphics*[#1]{#2}
    \caption{\label{#2} #4}
\end{center}
\begin{center}
    \includegraphics*[#1]{#3}
    \caption{\label{#3} #5}
\end{center}
\end{multicols}
\end{figure}
}

\makeatletter
\def\@citex[#1]#2{\immediate\write\@auxout{\string\citation{#2}}
\def\@citea{}\@cite{\@for\@citeb:=#2\do
{\@citea\def\@citea{; }\@ifundefined
{b@\@citeb}{{\bf ?}\@warning
{Citation `\@citeb' on page \thepage \space undefined}}%
{\csname b@\@citeb\endcsname}}}{#1}}
\makeatother

\usepackage[colorlinks,bookmarksopen,bookmarksnumbered,citecolor=red,urlcolor=red]{hyperref}
\hypersetup{pdftitle={Global sensitivity analysis with 2D hydraulic codes: application on
uncertainties related to high resolution topographic data},
bookmarks=true,
pdftoolbar=true,
pdfmenubar=true,
pdfauthor={M. Abily, O. Delestre, P. Gourbesville, N. Bertrand, C.-M. Duluc, Y. Richet},
pdfsubject={Uncertainty, flood hazard modelling, photogrammetry, 2D numerical modelling, global sensitivity analysis, 2D shallow water equations, urban flooding, Sobol index},
pdfcreator={Delestre},
pdfproducer={Delestre},
pdfkeywords={uncertainty} {flood hazard} {Shallow-Water equation} {Saint-Venant} {global sensitivity analysis} {Sobol index}
{high resolution modelling} {DSM} {surface model} {photogrammetry} {urban flooding} {2D numerical modelling}
 {FullSWOF} {Var river} {finite volume} {well-balanced} {hydrostatic reconstruction}}

\title{Global sensitivity analysis with 2D hydraulic codes: application on uncertainties related
to high resolution topographic data}

\author{{M. Abily}\footnote{Polytech'Nice Sophia \& URE Innovative-CiTy, University of Nice Sophia Antipolis,
France, e-mail : abily@polytech.unice.fr}, {O. Delestre}\footnote{Lab. J.A. Dieudonn\'e \& EPU Nice
Sophia, University of Nice, France, e-mail : delestre@math.unice.fr}, P. Gourbesville\footnote{Polytech'Nice
Sophia \& URE Innovative-CiTy, University of Nice Sophia Antipolis, France}, {N. Bertrand}\footnote{Institut
de Radioprotection et de S\^uret\'e Nucl\'eaire
(IRSN), PRP-DGE, SCAN, BEHRIG, France, e-mail : nathalie.bertrand@irsn.fr},
\\ 
{C.-M. Duluc}\footnote{Institut de Radioprotection et de S\^uret\'e Nucl\'eaire
(IRSN), PRP-DGE, SCAN, BEHRIG}\; and {Y. Richet}\footnote{Institut de Radioprotection
et de S\^uret\'e Nucl\'eaire (IRSN), PSN-EXP, SNC, France}}

\begin{document}
\maketitle

\begin{abstract}
Technologies such as aerial photogrammetry allow production of 3D topographic data including complex
environments such as urban areas. Therefore, it is possible to create High Resolution (HR) Digital Elevation
Models (DEM) incorporating thin above ground elements influencing overland flow paths. Even though
this category of “big data” has a high level of accuracy, there are still errors in measurements and
hypothesis under DEM elaboration. Moreover, operators look for optimizing spatial discretization
resolution in order to improve flood models computation time. Errors in measurement, errors in DEM
generation, and operator choices for inclusion of this data within 2D hydraulic model, might influence
results of flood models simulations. These errors and hypothesis may influence significantly flood
modelling results variability. The purpose of this study is to investigate uncertainties related to
($i$) the own error of high resolution topographic data, and ($ii$) the modeller choices when including
topographic data in hydraulic codes. The aim is to perform a Global Sensitivity Analysis (GSA) which
goes through a Monte-Carlo uncertainty propagation, to quantify impact of uncertainties, followed by a Sobol'
indices computation, to rank influence of identified parameters on result variability. A process using
a coupling of an environment for parametric computation (Prom\'eth\'ee) and a code relying on 2D
shallow water equations (FullSWOF\_2D) has been developed (P-FS tool). The study has been performed over
the lower part of the Var river valley using the estimated hydrograph of $1994$ flood event. HR topographic
data has been made available for the study area, which is $17.5\;\text{km}^2$, by Nice municipality.
Three uncertain parameters were studied: the measurement error (var. $E$), the level of details of
above-ground element representation in DEM (buildings, sidewalks, {\it etc.}) (var. $S$), and the
spatial discretization resolution (grid cell size for regular mesh) (var. $R$). Parameter
var. $E$ follows a probability density function, whereas parameters var. $S$ and var. $R$. are
discrete operator choices. Combining these parameters, a database of $2,000$ simulations has been
produced using P-FS tool implemented on a high performance computing structure. In our study case,
the output of interest is the maximal water surface reached during simulations. A stochastic sampling
on the produced result database has allowed to perform a Monte-Carlo approach. Sensitivity index have
been produced at given points of interest, enhancing the relative weight of each uncertain parameter
on variability of calculated overland flow. Perspectives for Sobol index maps production are put to
the light.
\end{abstract}

\section{Introduction}\label{sec:intro}
To understand or predict surface flow properties during an extreme food event, models based on 2D
Shallow Water Equations (SWEs) using high resolution description of the environment are commonly used
in practical engineering applications. In that case, the main role of hydraulic models is to finely
describe overland flow maximal water depth reached at some specific points or area of interest. In complex
urban environment, above ground surface features have a major influence on overland flow path, their
implementations (buildings, walls, sidewalks) in hydraulic model are therefore required as shown in
\cite{Abily13b,Abily13}. The representation of details surface features within models can be achieved through the use of
High Resolution Digital Elevation Models (HR DEMs).
\newline
Geomatics community intensively uses urban reconstruction relying on airborne topographic data gathering
technologies such as imagery and Light Detection and Ranging (LiDAR) scans to produce HR DEM \cite{Musialski13}. These
technologies allow producing DEM with a high accuracy level \cite{Lafarge10,Lafarge11,Mastin09}. Moreover, modern technologies,
such as Unmanned Aerial Vehicle (UVA) use, make high resolution LiDAR or imagery born data easily affordable
in terms of time and financial investments \cite{Remondino11,Nex13}. Consequently, hydraulic numerical modelling community
increasingly uses HR DEM information from airborne technologies to model urban flood \cite{Tsubaki10}. Among HR
topographic data, photogrammetry technology allows the production of 3D classified topographic data
\cite{Andres12}. This type of data is useful for surface hydraulic modelling community as it provides classified
information on complex environments. It gives the possibility to select useful information for a DEM
creation specifically adapted for flood modelling purposes \cite{Abily13}.
\newline
Even though HR classified data have high horizontal and vertical accuracy levels (in a range of few
centimeters), this data set is assorted of errors and uncertainties. Moreover, in order to optimize
models creation and numerical computation, hydraulic modellers make choices regarding procedure for this
type of dataset use. These sources of uncertainties might produce variability in hydraulic flood models
outputs. Addressing models output variability related to model input parameters uncertainty is an active
topic which is one of the main concern for practitioners and decision makers involved in assessment and
development of flood mitigation strategies \cite{ASN13}. To tackle a part of the uncertainty in modelling
approaches, practitioners are developing methods which enable to understand and reduce results variability
related to input parameters uncertainty such as Global Sensitivity Analysis (GSA) \cite{Herman13,Iooss11}. A GSA aims to
quantify the output uncertainty in the input factors, given by their uncertainty range and distribution
\cite{Nguyen15}. To do so, the deterministic code (2D hydraulic code in our case) is considered as a black box model
as described in \cite{Iooss11}:
\begin{equation}
 \begin{array}{rl}
  f: & \mathbb{R}^p \rightarrow \mathbb{R}\\
    &  X \mapsto Y=f(X)
 \end{array}\label{eq:Sobol}
\end{equation}
where $f$ is the model function, $X = (X_1;...;X_p)$ are $p$ independent input random variables with known
distribution and $Y$ is the output random variable. The principle of GSA method relies on estimation of
inputs variables variance contribution to output variance. An unique functional analysis of variance
(ANOVA) decomposition of any integral function into a sum of elementary functions allows to define the
sensitivity indices as explained in \cite{Iooss11}. Sobol's indices are defined as follow:
\begin{equation}
 S_i=Var\left[ \mathbb{E}(Y|X_i)\right]/Var(Y).
\end{equation}
First-order Sobol index indicates the contribution to the output variance of the main effect of each
input parameters. The production of Sobol index spatial distribution map is promising. Moreover, such maps
have been done in other application fields such as hydrology, hydrogeology and flood risk cost estimation
\cite{SaintGeours12}. GSA process most generally goes through uncertain parameters definition, uncertainty propagation
and results variability study. Such type of approach has been applied at an operational level in 1D
hydraulic modelling studies by public institutions and consulting companies \cite{Nguyen15}. For 2D free surface
modelling, GSA approach is still at an exploratory level. Indeed, GSA requires application of a specific
protocol and development of adapted tools. Moreover, it requires important computational resources. 
\newline
The purpose of this study is to investigate on uncertainties related to HR topographic data use for
hydraulic modelling. Two categories of uncertain parameters are considered in our approach: the first
category is inherent to HR topographic data internal errors (measurement errors) and the second category
is related to operator choices for this type of data inclusion in 2D hydraulic codes.
\newline
This paper presents the results of an applied GSA approach performed over a 2D flood river event
modelling case. The aim of our study is to rank the impact of uncertainties related to HR topographic
use. To achieve this aim, a protocol and a tool for GSA application to 2D hydraulic codes have been
developed. Input parameters considered as introducing uncertainty are chosen and a probability distribution
is attributed to each uncertain input parameters. A Monte-Carlo uncertainty propagation is then carried
out, uncertainties are quantified and the influence of selected input parameters are ranked by the
computation of the Sobol indices. Section \ref{sec:mat-meth} introduces the data and methodology used and followed.
Then, section \ref{sec:Results and discussion} presents the first results, main outcomes and perspectives.

\section{Material and methods}\label{sec:mat-meth}

\subsection{Flood event scenario}\label{subsec:Flood-event-scenarioh}

The $5^{th}$ of November $1994$, an intense rainfall event occurred in the Var catchment (France),
leading to serious flooding in the low Var river valley \cite{Lavabre96}. For our study, hydraulic conditions
of this historical event were used as a framework for a test scenario. The study area has been
restricted to the last five kilometers of the low Var valley. Since $1994$, the urban area has changed a
lot, as levees, dikes and urban structures have been intensively constructed and it has to be reminded
that the objective here was not to reproduce the flood event itself. For the GSA approach, the hydraulic
parameters of the model are set identically for the simulations (as described below), only the input DEM
changes from one simulation to another following the strategy defined in next section.
\newline
The 2D hydraulic code is FullSWOF\_2D \cite{Delestre14,Delestre14b}. FullSWOF\_2D relies on 2D SWEs and uses a finite
volume approach over a regular Cartesian grid. An estimated hydrograph of the $5^{th}$ of November
$1994$ flood event as described in \cite{Guinot03} is used as the upstream boundary condition of the low Var
river valley. To shorten the simulation length, we chose to run a constant $1,500\;\text{m}^3.\text{s}^{-1}$
discharge for $3$ hours, to reach a steady state with a water level in the riverbed just half a meter
below the elevation of the flood plain. The reached steady state is used as an initial condition
(or hot start) for the other simulations. For the GSA simulations the unit hydrograph is then run until
the estimated peak discharge ($3,700\;\text{m}^3.\text{s}^{-1}$) and decreases until a significant
diminution of the overland flow water depth is observed. The Manning's friction coefficient $n$ is
spatially uniform on overland flow areas with a standard value of $0.015$ which corresponds to a
concrete surfacing. No energy loss properties have been included in the 2D hydraulic model to represent
the bridges, piers or weirs. Downstream boundary condition is an open sea level with a Neumann
boundary condition. 
\newline
For our application, the 3D classified data of the low Var river valley is used to generate specific
DEM adapted to surface hydraulic modelling.

\subsection{Photo-interpreted high resolution topographic data of the low Var valley}\label{subsec:Photo-interpreted}

Aerial Photogrammetry technology allows to measure 3D coordinates of a surface and its features using
2D pictures taken from different positions. The overlapping between pictures allows calculating, through
an aerotriangulation calculation step, 3D properties of space and features based on stereoscopy
principle \cite{Egels04,Lu07}. Photo-interpretation allows creation of vectorial information based on
photogrammetric dataset. The 3D classification of features based on photo-interpretation allows getting
3D high resolution topographic data over a territory offering large and adaptable perspectives for its
exploitation for different purposes \cite{Andres12}. A photo-interpreted dataset is composed of classes of points,
polylines and polygons digitalized based on photogrammetric data. Important aspects in the
photo-interpretation process are classes' definition, dataset quality and techniques used for
photo-interpretation. Both will impact the design of the output classified dataset \cite{Lu07}. 
\newline
A HR photogrammetric 3D classified data gathering campaign has been held in $2010-2011$ covering
$400\;\text{km}^2$ of Nice municipality \cite{Andres12}. Aerial pictures have a pixel resolution of $0.1\;\text{m}$ at the
ground level. Features have been photo-interpreted by human operators under vectorial form in
$50$ different classes. These classes of elements include large above ground features such as
building, roads, bridges, sidewalks, {\it etc.}. Thin above ground features (like concrete walls,
road-gutters, stairs, {\it etc.}) are included in classes. An important number of georeferencing
markers were used (about $200$). Globally, over the whole spatial extent of the data gathering campaign,
mean accuracy of the classified data is $0.3\;\text{m}$ and $0.25\;\text{m}$ respectively in horizontal and vertical
dimension. For the low Var river valley area, a low flight elevation combined with a high level of
overlapping among aerial pictures ($80 \%$), have conducted to a higher level of accuracy. In the low
Var river valley sector, classified data mean horizontal and vertical mean accuracy is $0.2\;\text{m}$. This mean
error value encompasses errors, due to material accuracy limits, to bias and to nuggets, which occurs
within the photogrammetric data. For this data set, errors in photo-interpretation are estimated to
represent $5\%$ of the total number of elements. This percentage of accuracy represents errors in
photo-interpretation, which results from feature misinterpretation, addition or omission. To
control and ensure both, average level of accuracy and level of errors in photo-interpretation,
the municipality has carried out a terrestrial control of data accuracy over $10\%$ of the domain covered
by the photogrammetric campaign. 

\subsection{Global Sensitivity Analysis}\label{subsec:GSA}

A GSA method quantifies the influence of uncertain input variables on the variability in numeric models
outputs. To implement a GSA approach, it is necessary ($i$) to identify inputs and assess their
probability distribution, ($ii$) to propagate uncertainty within the model ({\it e.g.} using a Monte
Carlo approach) and ($iii$) to rank the effects of input variability on the output variability through
functional variance decomposition method such as calculation of Sobol indices (eq. \ref{eq:Sobol}).

\subsubsection{Uncertain input parameters}\label{subsec:Uncertain-input-parameters}

To encompass uncertainty related to measurement error in HR topographic dataset, Var. $E$ is considered.
Two other uncertain parameters are also considered as modeller choices: the level of details of
above-ground elements included in DEM (Var. $S$), and the spatial discretization resolution (Var $R$).
Parameters Var. $S$, $E$ and $R$ are independent parameters considered as described below.
\newline
{\bf Var. $E$: measurement errors of HR topographic dataset --} In each cell of the DEM having the
finest resolution ($1\;\text{m}$), this parameter introduces a random error. For our study, only the altimetry
errors are taken into account as the planimetric dimension of the error is assumed to be relatively
less significant for hydraulic study purpose compared to altimetry error. This altimetry measurement
error follows a Gaussian probability density function $\mathcal{N} (0; 0.2)$, where the standard deviation is
equal to the mean global error value ($0.2\;\text{m}$). This error introduction is spatially homogeneous.
This approach is a first approximation: mean error could be spatialized in different sub-areas having
physical properties, which would impact spatial patterns of error value. Moreover, errors in
photo-interpretation (classification) are not considered in our study. One hundred grids of random
errors are generated and named $E1$ to $E100$.
\newline
{\bf Var. $S$: modeller choices for DEM creation --} This parameter represents modeller choices for DEM
creation, taking advantage of selection possibilities offered by above described classified topographic
data. Four discrete schemes are considered: ($i$) $S1$, is the DTM of the study case, ($ii$) $S2$, the
elevation information of buildings added to $S1$, ($iii$) $S3$ the elevation information of walls added to
$S2$, and ($iv$) $S4$, elevation information of concrete features in streets added to $S3$. Var. $S$
parameter is included in the SA as a categorical ordinal parameter. These discrete modeller choices
are considered as having the same probability. Four DEMs are generated at resolution $1\;\text{m}$, $S1$ to $S4$.
\newline
{\bf Var. $R$: modeller choices for mesh spatial resolution --} When included in 2D models, HR DEM
information is spatially and temporally discretized. FullSWOF is based on structured mesh, therefore
the DEM grid can be directly included as a computational grid without effort for mesh creation.
Nevertheless, for practical application, optimization of computational time/accuracy ratio often
goes through a mesh degradation process when a HR DEM is used. Var. $R$ represents modeller choices,
when decreasing regular mesh resolution. Var. $R$ parameter takes $5$ discrete values : $1$, $2$, $3$, $4$ or $5\;\text{m}$.

\subsubsection{Applied protocol for uncertainty propagation}\label{subsec:Applied protocol}

\begin{figure}[htbp]
\begin{center}
\includegraphics[width=0.92\textwidth]{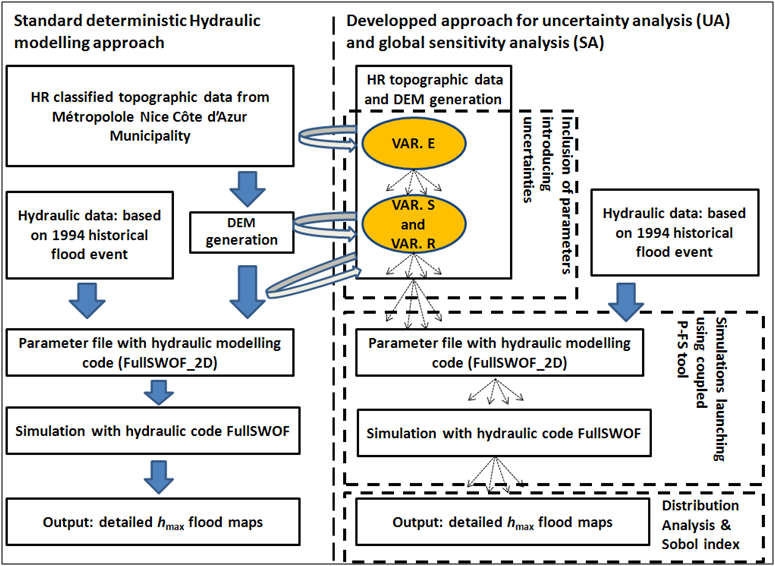}
\caption{Implemented GSA approach.}
\label{fig1}
\end{center}
\end{figure}

To create the HR DEMs, the following approach has been carried out. A HR DTM using multiple ground
level information sources (points, polygons and polylines) is created and provided at a $0.5\;\text{m}$ resolution
by The GIS Department of Nice C\^ote d'Azur Metropolis (DIGNCA). The HR DEM resolution is degraded
to $1\;\text{m}$ resolution. At this resolution the number of mesh cells is above $17.8$ million. Then, a selection
procedure among classified data is performed. This selection is achieved by considering concrete
elements which can influence overland flow drainage path only. It includes dikes, buildings, walls
and "concrete" above ground elements (such as sidewalks, road gutters, roundabout, doors steps,
{\it etc.}). $12$ classes are selected among the $50$ classes of the 3D photo-interpreted dataset. During
this step, polylines giving information on elevated roads and bridges, which might block overland flow
paths, are removed. The remaining total number of polylines is $52,600$. Selected above-ground features are
aggregated in $3$ groups of features (buildings, walls and concrete street features). Extruding elevation
information of selected polylines groups on the DTM ($S1$), four $1\;\text{m}$ resolution DEMs, $S1$ to $S4$,
are produced. The previously described method has allowed inclusion of thin elements impacting
flow behavior of infra-metric dimension, oversized to metric size, in the $1\;\text{m}$ resolution regular
mesh. Then, $100$ grids of var. $E$ are produced and added to var. $S1$, $S2$, $S3$ and $S4$
at resolution $1\;\text{m}$. These $400$ DEMs are used to create $2,000$ DEMs with a resolution from $1$ to $5\;\text{m}$.
DEMs are named $SmRnEx$, with the parameters $m$ between $\left[1; 4\right]$, $n$ between $\left[1; 5\right]$ and $x$ between
$\left[1; 100\right]$. These DEMs are used in the coupled parametric environment (Prom\'eth\'ee) -- 2D hydraulic
code (FullSWOF\_2D) through parameterization of the input file, and integrated in whole GSA as summed
up in Figure \ref{fig1}. Prom\'eth\'ee-FullSWOF (P-FS) is presented in more detail in next section

\subsubsection{Operational tool and setup developed
for uncertainty analysis and Sobol index mapping}\label{subsec:Operational tool and setup}

To apply a GSA with 2D Hydraulic models, a coupling between Prom\'eth\'ee a code allowing a parametric
environment of other codes, has been performed with FullSWOF\_2D, a two-dimensional SWE based
hydraulic code. The coupling procedure has taken advantage of previous coupling experience of
Prom\'eth\'ee with 1D SWE based hydraulic code \cite{Nguyen15}. The coupled code Prom\'eth\'ee-FullSWOF
(P-FS) has been performed on a HPC computation structure. The aim was to use tool for a proof of
concept of protocol application requiring extensive computational resources.
\newline

{\bf FullSWOF\_2D --} FullSWOF\_2D (Full Shallow Water equation for Overland Flow in 2 dimensions)
is a code developed as a free software based on 2D SWE \cite{Delestre14}.  In FullSWOF\_2D, the 2D SWE are
solved thanks to a well-balanced finite volume scheme based on the hydrostatic reconstruction.
The finite volume scheme, which is suited for a system of conservation low, is applied on a structured
spatial discretization, using regular Cartesian mesh. For the temporal discretization, a variable time
step is used based on the CFL criterion. The hydrostatic reconstruction (which is a well-balanced
numerical strategy) allows to ensure that the numerical treatment of the system preserves water
depth positivity and does not create numerical oscillation in case of a steady states, where pressures
in the flux are balanced with the source term here (topography). Different solvers can be used
HLL, Rusanov, Kinetic, VFROE combined with first order or second order (MUSCL or ENO) reconstruction.
FullSWOF\_2D is an object oriented software developed in C++. Two parallel versions of the code have
been developed allowing to run calculations under HPC structures \cite{Cordier13}.
\newline
{\bf Prom\'eth\'ee-FullSWOF --} Prom\'eth\'ee software is coupled with FullSWOF\_2D. Prom\'eth\'ee
is an environment for parametric computation, allowing to carry out uncertainties propagation
study, when coupled to a code. This software is an open source environment developed by IRSN
(\url{http://promethee.irsn.org/doku.php}). The main interest of Prom\'eth\'ee is the fact that it
allows the parameterization of any numerical code. Also, it is optimized for intensive computing
resources use. Moreover, statistical post-treatment can be performed using Prom\'eth\'ee as it
integrates R statistical computing environment \cite{Ihaka98}. The coupled code Prom\'eth\'ee/FullSWOF (P-FS)
is used to automatically launch parameterized computation through R interface under Linux OS. A graphic
user interface is available under Windows OS, but in case of large number of simulation launching,
the use of this OS has shown limitations as described in \cite{Nguyen15}. A maximum of $30$ calculations can be
run simultaneously, with the use of $30$ “daemons”.

Once simulations were completed with P-FS, GSA analysis has been carried out. First, the convergence
of the results, in other word it has been checked that the number of simulations is large enough to
generate a representative sample of the uncertainties associated to the studied source. Then 
uncertainties analysis was conducted, followed by the calculation of Sobol indices. The selected
output of interest is the overland flow water surface elevation ($h_{max}+z$).

\section{Results and discussion}\label{sec:Results and discussion}

\subsection{Operational achievement of the approach}\label{subsec:Operational achievement of the approach}

A performed version of P-FS couple allows to run simulations with selected set of input parameters
(Var. $E$, $S$ and $R$). The coupled tool is operational on the M\'esocentre HPC computation center
and P-FS would be transposable over any common high performance computation cluster, requiring only
slight changes in the coupling part of the codes. Through the use of R commands, it is possible
to launch several calculations. Using “Daemons”, up to $30$ simulations can be launched at a time.
The calculations running time of our simulation is significant. Indeed, this computation time is
highly dependent of mesh resolution as the $dx$ will directly impact the CFL dependent $dt$.
Over a $12$ cores node of the M\'esocentre HPC, the computation time is $2$, $6$, $12$, $24$, $80$ hours
respectively for $5$, $4$, $3$, $2$, $1\;\text{m}$ resolution grids. Using about $400,000$ CPU hours, it has been
possible to run $1,500$ simulations. Out of these $1,500$ simulations, few runs (about $30$) have shown
numerical errors leading to computational crash. These simulations have been removed from our set
of simulations used to carry out the GSA. This will be clarified in future work, but errors are
possibly due to numerical instabilities generated by important topographic gradient change at the
boundary condition. This first data set of output allows us to carry out a first UA and GSA. The
remaining $500$ simulations are mainly for $R1$ and $R2$ resolutions, which are the most
resource-demanding simulations, and will be run in a close future using more than one HPC node per each
run to decrease running time.

\begin{figure}[htbp]
\begin{center}
\includegraphics[width=0.98\textwidth]{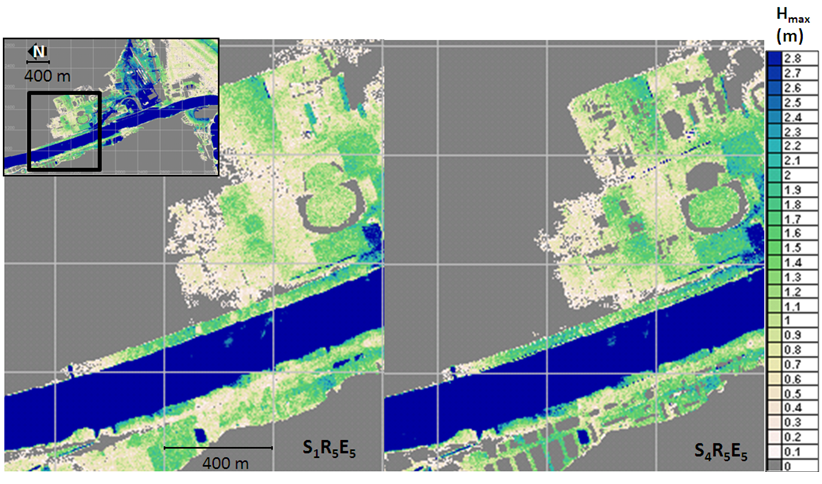}
\caption{Illustration of $h_{max}$ value for two given simulations where Var. $S$ changes.}
\label{fig2}
\end{center}
\end{figure}

The variable of interest is the maximal water surface elevation ($h_{max}+z$) reached during a given simulation
at different locations, with $h_{max}$ the maximal water depth reached at the point of interest and $z$
the DEM surface elevation at this point. The figure \ref{fig2} illustrates the difference of $h_{max}+z$
obtained between two simulations when Var. $S$ varies. At different points of interest (see next section)
difference in $h_{max}+z$ value can be significant. The analyses carried out over $40$ points of interests,
highlight the influence of the selected variables. For example, for a given scenario when only Var. $E$
varies, an up to $0.5\;\text{m}$ difference in $h_{max}+z$ can be observed. Between all the $1,500$ scenarios a
maximal difference of $1.26\;\text{m}$ in $h_{max}+z$ estimation at one of the point of interest is observed.

\subsection{Local results for a point of interest}\label{subsec:Local results for a point of interest}

In the first place, a local analysis of the influence of topographic parameters has been achieved; $40$ points
of interest have been selected and used for the analyses. 
\newline

\begin{figure}[htbp]
\begin{center}
\includegraphics[width=0.98\textwidth]{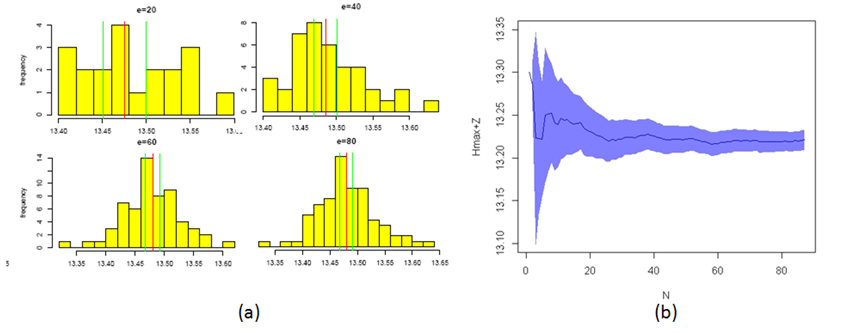}
\caption{(a): Illustration of $h_{max}+z$ distribution and (b) convergence of mean and CI with fixed
Var. $S$ and Var. $R$ when increasing sample of size of Var. $E$.}
\label{fig3}
\end{center}
\end{figure}

The convergence for Var. $E$ has been analyzed for the different points. As illustrated by figure
\ref{fig3}(a), it appears that the distribution and the standard deviations of $h_{max}+z$,
become stable with a sample size ($N$) of Var. $E$ is around $40$ to $50$. This gives qualitatively a
first idea of what should be the minimum size of sample $N$ of Var. $E$ to allow performing reliable
statistical analysis with an acceptable level of convergence.
\newline

To strengthen these findings, tests of convergence have been performed observing the evolution of mean
$h_{max}+z$ value and the $95\%$ confidence interval (CI) when $N$ size increases. Figure \ref{fig3}(b)
shows this result for a given point of interest. The analysis shows the sample size is above $30$,
the mean and the CI become stable. It has to be noticed that similar results are obtained with the
other selected points of interest, $30$ to $40$ realizations are sufficient to generate a representative
sample of the uncertainties associated to the Var. $E$.

\begin{figure}[htbp]
\begin{center}
\includegraphics[width=0.98\textwidth]{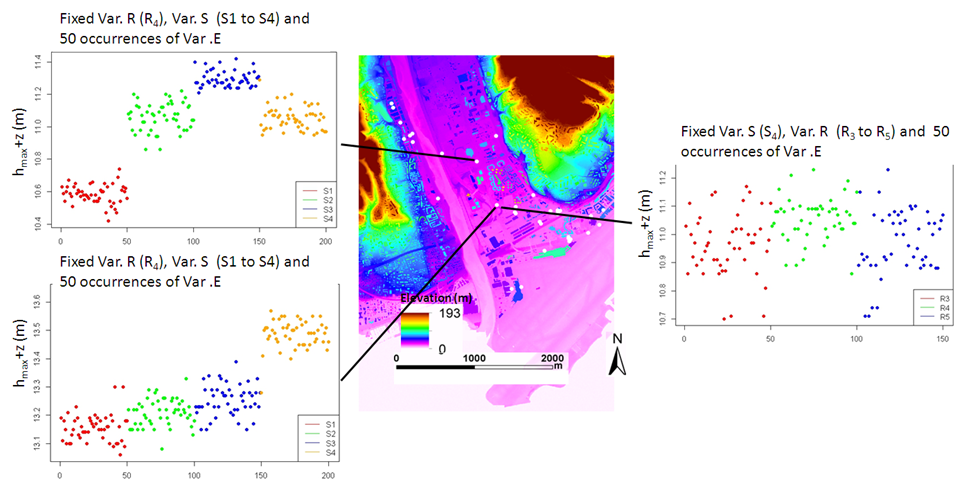}
\caption{Output ($h_{max}+z$) distribution plots at two points of interest with one fixed value (either
Var. $S$ or Var. $R$).}
\label{fig4}
\end{center}
\end{figure}

When looking at the output variable of interest $h_{max}+z$, it is relevant to check its distribution
behavior for a fixed value of one of the two discrete input parameters (Var. $R$ or Var. $S$). This
has been done with a subset of sample $N$ Var. $E$ equal to 50 for each discrete value of non fixed
variables Var. $R$ and Var. $S$ (figure \ref{fig4}). This approach helps to make a qualitative description
of the output distribution behavior relatively to the non fixed parameters. This test has been carried
out for $40$ different points of interest. Figure \ref{fig4} illustrates the main observations which
can be effectuated using different distribution plots for fixed Var. $R$ or for fixed value of Var. $S$.
Results show that for a given value of Var. $R$, Var. $E$ impact over variability of $h_{max}+z$
is relatively less significant than the impact of Var. $S$ (discrete choices). It has also been
observed that increasing the level of geometric details included in DEM (Var. $S$) will not involve
linear variations in $h_{max}+z$ values. Indeed, detailed above ground feature implementation lead to
more local effects in terms of overland flow path modification and consequently, highly impact local
$h_{max}+z$ values. When focusing on $h_{max}+z$ distribution for varying discrete mesh resolution
(Var. $R$) for a given fixed Var. $S$ value (figure \ref{fig4}), it is observed that, comparatively
to effect of Var. $E$ in output distribution, the influence generated by Var. $R$ on $h_{max}+z$
values is negligible. This finding does not consider the finest Var. $R$ ($1\;\text{m}$) as less than $50$
realizations of Var. $E$ were available so far with these resolutions.

\begin{figure}[htbp]
\begin{center}
\includegraphics[width=0.98\textwidth]{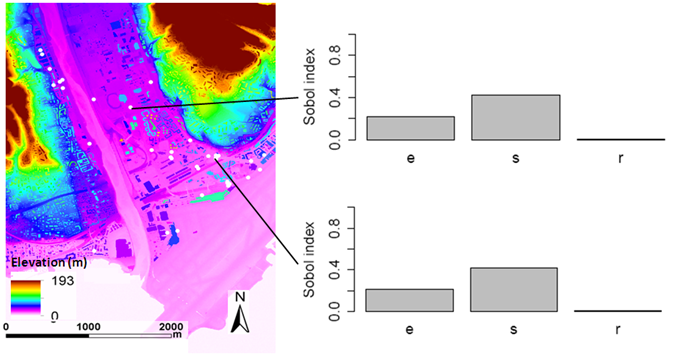}
\caption{First-order Sobol indices for two selected points.}
\label{fig5}
\end{center}
\end{figure}

The final step of the GSA approach calculates the Sobol indices (Figure \ref{fig5}). As mentioned
previously Var. $R1$ has not been considered for the calculation. The parameter which influences the
most $h_{max}+z$ is Var. $S$. Concerning the Sobol indices, it has to be mentioned that the sum of
Sobol indices should be one, in our case the sum is smaller than one. Similar results have been
highlighted by \cite{Iooss11}. This finding is due to the parameters cross variation. Moreover the inclusion
of results at the highest resolution ($1\;\text{m}$) might increase the cross variation effects. These promising
results for analysis are already useful and further analysis to observe cross variation effects as well
as spatial variation of Sobol index are in progress. So far it can be observed that $32$ of the $40$ points
of interest have Var. $S$ with the highest Sobol index and $8$ points have Var. $R$ with the highest Sobol
index. These two parameters are modeller choices. In the $32$ cases where Var. $S$ has the highest
Sobol Index, $50\%$ of the points have Var. $R$ in the second rank and $50\%$ Var. $E$. 

\subsection{Perspectives and analyze for further work}\label{subsec:Perspectives and analyze for further work}

\begin{figure}[htbp]
\begin{center}
\includegraphics[width=0.98\textwidth]{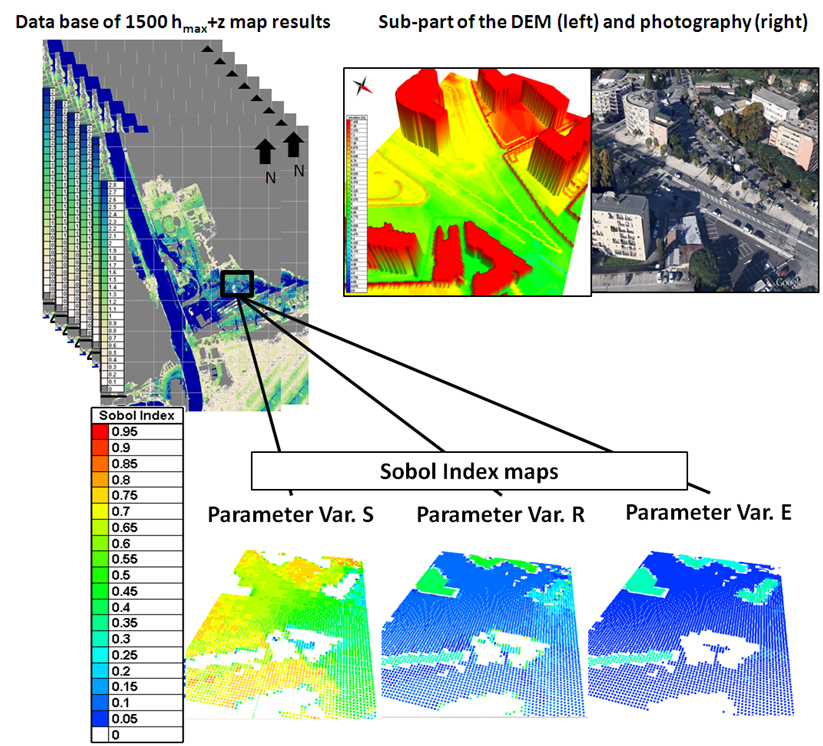}
\caption{Sobol index maps over $150\;\text{m}$ long per $200\;\text{m}$ large part of the domain.}
\label{fig6}
\end{center}
\end{figure}

One of the main advantages of two-dimensional hydraulic models is their spatial distribution over the
area modeled. Therefore, uncertainties related to topography variability can be spatially represented
for the Var river valley. Sobol index maps are presented in figure \ref{fig6} over a sub-area. In a near future
and integrate cross variation effect.

Nevertheless, as it has been mentioned, our representation of Var. $E$, using a spatially uniform
distribution function of average measurement error is a first simple approximation. Indeed, the
average error is a spatially varying function of physical properties (slope notably). Assigning
differents spatially varying parameters of Var. $E$ distribution function would be a more sophisticated
approach. Moreover, errors in photo-interpretation, which will spatially impact Var. $S$, deserve
to be investigated as well.

\section{Conclusions}\label{sec:conclusions}

In this study, a GSA is performed based on Sobol sensitivity analysis to quantify the uncertainties
related to high resolution topography data inclusion in two-dimensional hydraulic model. Input parameters
considered in the study encompass both measurement errors, and modeller choices for high resolution
topographic data used in models. The applied GSA method relies ($i$) on the use of a specific tool
coupling a parametric environment with a 2D hydraulic modelling tool Prom\'eth\'ee-FullSWOF\_2D,
and ($ii$) on the use of high performance computation resources. The implemented approach is able to
highlight the uncertainties generated by topography parameters and operator choices when including
high resolution data in hydraulic models.
\newline

$1,500$ simulations have been effectuated at this stage of the study. Convergence of the approach is
checked and output distribution analyzed for qualitative apprehension of input parameters local effects
on maximal computed overland flow ($h_{max}+z$).  The major source of uncertainty related to water
elevation $h_{max}+z$ is the modeller choice Var. $S$ which is the choice of operator regarding above
ground features included in DEM used for hydraulic simulation. Errors related to measurement Var. $E$
significantly impact variability of $h_{max}+z$ model outputs but in a relatively smaller extent.
Regarding Var. $R$, further investigations will be undertaken when more realizations will be
available. Overall, these results confirm the relative importance of the uncertainty in topography
input data. The main limits of the approach are concerning the way Var. $E$ and Var. $S$ are integrated
in the analysis. Future work will be focused on the design of Sobol map index over the whole flood extent.

\section*{Acknowledgments}

Photogrammetric and photo-interpreted dataset used for this study have been kindly provided by Nice
C\^ote d'Azur Metropolis for research purpose. Technical expertise on DIGNCA dataset has been provided
by G. Tacet and F. Largeron. This work was granted access to the HPC resources of Aix-Marseille
Universit\'e financed by the project Equip@Meso (ANR-10-EQPX-29-01) of the program "Investissements d'Avenir"
supervised by the Agence Nationale pour la Recherche. Technical support for codes adaptation on high
performance computation centers has been provided by T. Nguyen, J. Brou, F. Lebas., H. Coullon and P. Navarro.


\end{document}